\documentclass[preprint,superscriptaddress,aps]{revtex4}
\usepackage{amsfonts}
\usepackage{amssymb}
\usepackage{amsmath}
\usepackage{graphics}
\usepackage{graphicx}
\usepackage{color}
\usepackage[T1]{fontenc}
\usepackage{hyperref}
\newcommand{\bea}{\begin{eqnarray}}
\newcommand{\eea}{\end{eqnarray}}

\begin{document}
\title{On the effective superpotential in the generic higher-derivative three-dimensional scalar superfield theory}

\author{F. S. Gama}
\email{fgama@fisica.ufpb.br}
\affiliation{Departamento de F\'{\i}sica, Universidade Federal da Para\'{\i}ba\\
 Caixa Postal 5008, 58051-970, Jo\~ao Pessoa, Para\'{\i}ba, Brazil}

\author{J. R. Nascimento}
\email{jroberto@fisica.ufpb.br}
\affiliation{Departamento de F\'{\i}sica, Universidade Federal da Para\'{\i}ba\\
 Caixa Postal 5008, 58051-970, Jo\~ao Pessoa, Para\'{\i}ba, Brazil}

\author{A. Yu. Petrov}
\email{petrov@fisica.ufpb.br}
\affiliation{Departamento de F\'{\i}sica, Universidade Federal da Para\'{\i}ba\\
 Caixa Postal 5008, 58051-970, Jo\~ao Pessoa, Para\'{\i}ba, Brazil}

\begin{abstract}
We formulate a generic three-dimensional higher-derivative superfield theory for self-interacting scalar superfield action. We consider the cases of real and complex scalar superfields. For these theories, we explicitly calculate the one-loop effective potential.
\end{abstract}

\maketitle

The higher-derivative field theory models have acquired great scientific attention. The initial motivation for their study was, certainly, the fact that for these theories the ultraviolet behaviour is essentially improved, which allows to treat introduction of higher derivatives as a possible solution for the problem of renormalizability of gravity \cite{Stelle}. Another important example of the higher-derivative theory is the Lee-Wick theory \cite{LW} representing itself as a higher-derivative extension of QED. Within the supersymmetry context, the higher-derivative terms arose first in the context of supergravity, as a consequence of superconformal anomalies \cite{BK}, and quantum properties of the resulting theory (which describes the higher-derivative dynamics of the chiral superfield playing the role of one of supergravity prepotentials) were considered in \cite{dilaton}. Further, the more generic examples of higher-derivative superfield four-dimensional theories were discussed in \cite{HD,HD1}, where the one-loop effective potential for these theories has been explicitly calculated.

At the same time, the three-dimensional superspace is known as a very convenient laboratory for study of different properties of field theory models. Therefore studies of the three-dimensional higher-derivative superfield theories seem to be very interesting. In our recent paper \cite{IYouHe}, the higher derivatives were introduced in the superfield gauge theories, while the matter sector was maintained to be the same. In this paper, we continue that study by introducing of higher derivatives into the matter sector. We study the one-loop effective action for a purely scalar higher-derivative theory. The results which we obtain are rather generic. Generic results are important due to the fact that they gather in a single equation information on a wide range of models, thus allowing that one can determine more clearly the influence that a set of models may have on a given physical quantity. Within this work, in particular, we determine what is the influence of the addition of higher-derivative terms in the three-dimensional scalar superfield theory have on the form of the one-loop effective potential. Within this paper we use the method based on summation over Feynman supergraphs, while, in the scalar superfield theory without higher derivatives, the functional approach based on the proper-time method was used \cite{ourEP}.

To start our study, let us recall the action of a general scalar superfield action without higher derivatives \cite{SGRS}:
\bea
\label{realfield}
S_{R}=\frac{1}{2}\int d^5z\Phi D^2\Phi+\int d^5zV(\Phi) \ ,
\eea
where $\Phi$ is a real scalar superfield and $V(\Phi)$ is the superpotential. In the usual case, without higher derivatives, the superpotential must be a polynomial of fourth  or less order for the renormalizability. However, in the higher-derivative case such a restriction is not necessary.

Now, let us try to generalize this theory introducing higher derivatives in the action (\ref{realfield}). Since there is an infinity of ways to accomplish this, we will demand, for simplicity, that the new action does not contain interaction terms with higher derivatives. The reason for this choice is that interaction terms with higher derivatives tend to have worse ultraviolet behaviour than those without ones. Consequently, we will consider in this work the higher derivative theory described by the action:
\bea
\label{higher1}
S_{HR}=\frac{1}{2}\int d^5z\Phi \hat R\Phi+\int d^5zV(\Phi) \ ,
\eea
where $\hat R$ is a some scalar operator which is a function of covariant derivatives and some constants. Due to the identity $(D^2)^2=\Box$, we can infer that $\hat R=g(\Box)+f(\Box)D^2$ is the most general choice for the scalar operator. Of course, when we take $g(\Box)=0$ and $f(\Box)=1$, we recover the model (\ref{realfield}).

Let us briefly discuss the renormalizability of this theory. First, since $\hat R=f(\Box)D^2+g(\Box)$, let us, for the sake of the concreteness suppose that the number of derivatives contributing to the superficial degree of divergence, in the term $f(\Box)D^2$ is no less that in the term $g(\Box)$ as it occurs in many cases (in the opposite case the situation does not differ essentially). Let us also suggest that the function $f(\Box)$ has the (higher) order $a$ in the space-time derivatives (the usual case is, of course, $a=0$). In this case, the superficial degree of divergence for this theory can be shown to be equal to $\omega=2+(1-a)P-2V$. Since we suggest that the function $f$ depends only on the d'Alembertian operator $\Box$ by the symmetry reasons, one must have $a\ge 2$, which implies the all-loop finiteness of the theory.

For a quantum system with a superfield $\phi$ interacting with a background superfield $\Phi$, we represent the effective action $\Gamma[\Phi]$ as a generating functional of the one-particle-irreducible vertex Green functions \cite{BO}:
\bea
\label{definitionEA}
\exp{(\frac{1}{\hbar}\Gamma[\Phi])}=\int {\cal D}\phi\exp{(\frac{1}{\hbar}S[\Phi+\sqrt{\hbar}\phi])}|_{1PI} \ .
\eea
The effective action encodes the full quantum dynamics of the theory. The usual methods of calculation of $\Gamma[\Phi]$ are peformed by means of perturbative series in powers of $\hbar$, the so-called loop expansion. Let us consider the application of relation
\bea
\label{loopexpansionEA}
\Gamma[\Phi]=S[\Phi]+\hbar\Gamma^{(1)}[\Phi]+\hbar^2\Gamma^{(2)}[\Phi]+\ldots \ ,
\eea
for (\ref{definitionEA}). In particular, the one-loop correction has the form
\bea
\label{one-loopEA}
\exp{\Gamma^{(1)}[\Phi]}=\int {\cal D}\phi\exp{(S_2[\Phi;\phi])} \ ,
\eea
where $S_2[\Phi;\phi]$ is the second-order part of the classical action functional $\frac{1}{\hbar}S[\Phi+\sqrt{\hbar}\phi]$ in quantum superfields.

The general structure of the effective action can be cast in a form \cite{BuKu,ourEP}
\bea
\label{derexpEA}
\Gamma[\Phi]=\int d^5z K(\Phi)+\int d^5z F(D_A\Phi,D_AD_B\Phi,D_AD_BD_C\Phi,\ldots;\Phi) \ ,
\eea
where $D_A=(D_\alpha,\partial_{\alpha\beta})$. The function $F(D_A\Phi,\ldots;\Phi)$ has the property that when all derivatives of the superfields are equal to zero, it vanishes identically. From (\ref{derexpEA}), we can define the K\"{a}hler effective potential (KEP) $K(\Phi)$ as the zero-order term in the covariant derivative expansion of the effective action of a background scalar superfield. The KEP carries all the quantum information about the slowly varying background superfields in superspace. Similarly to the effective action, KEP is also calculated by means of the loop expansion
\bea
K(\Phi)=V(\Phi)+\hbar K^{(1)}(\Phi)+\hbar^2 K^{(2)}(\Phi)+\ldots \ .
\eea
In this work, we are only interested in the calculation of the correction $K^{(1)}(\Phi)$. The technique that we will use in order to calculate such object will be the
Feynman supergraph technique which was examined in \cite{SYM}.  From (\ref{one-loopEA}), it is convenient to derive the propagators and vertices from the functional $S_2[\Phi;\phi]$. Therefore, expanding (\ref{higher1}) around a background superfield $\Phi+\sqrt{\hbar}\phi$ and keeping only the quadratic term in quantum fluctuation $\phi$, we get
\bea
\label{quadraticfields}
S_2[\Phi;\phi]=\frac{1}{2}\int d^5z\phi(g(\Box)+f(\Box)D^2)\phi+\frac{1}{2}\int d^5zV^{\prime\prime}(\Phi)\phi^2 \ ,
\eea
where we use a shorthand notation $V^{\prime\prime}(\Phi)=\frac{\partial^2V(\Phi)}{\partial\Phi^2}$. By convenience, propagators can be defined from the terms that are independent of $\Phi$, and vertices can be defined from the ones in which the $\phi$ interacts with $\Phi$. Hence, from (\ref{quadraticfields}), it follows that the propagator is given by
\bea
\label{propagator1}
\langle\phi(1)\phi(2)\rangle=-\frac{1}{g(k^2)+f(k^2)D_1^2}\delta_{12} \ ,
\eea
where $\delta_{12}=\delta^2(\theta_1-\theta_2)$ is an usual Grassmannian delta function.

In order to calculate the one-loop correction to the KEP, we will proceed in three steps. First, we draw all one-loop supergraphs allowed by (\ref{quadraticfields}). Second, we discard supergraphs involving covariant derivatives of $\Phi$ and calculate the contributions of each supergraph, with the external momenta equal zero, to the effective action. Last, we sum all contributions and calculate the integral over the momenta. The result will be just the KEP.

Let us start the calculations of the one-loop supergraphs, that is, those ones involving the scalar superfield propagators (\ref{propagator1}) connecting the vertices $V^{\prime\prime}(\Phi)\phi^2$. Such supergraphs exhibit structures given at Fig. 1.

\begin{figure}[!h]
\begin{center}
\includegraphics[angle=0,scale=0.40]{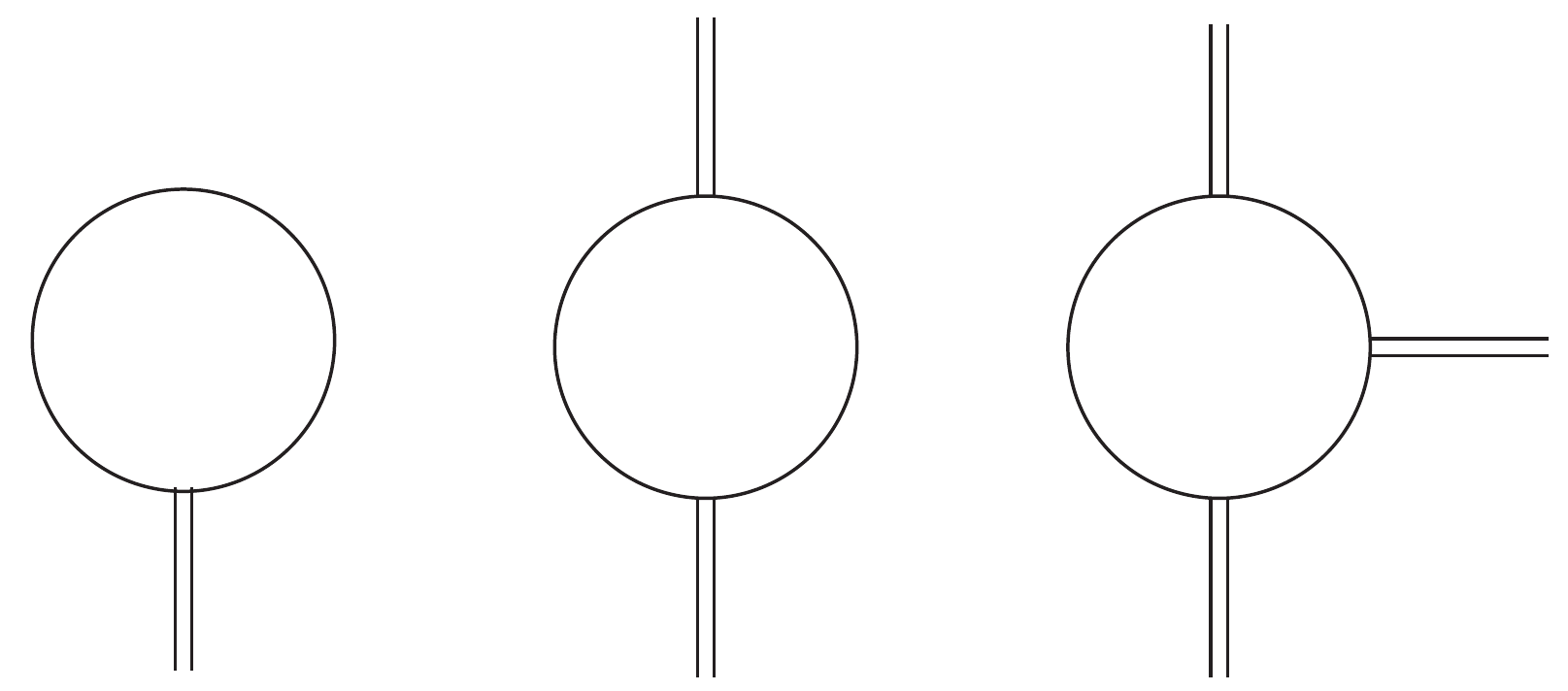}
\end{center}
\caption{One-loop supergraphs.}
\end{figure}

We can compute all the contributions by noting that each supergraph above is formed by $n$ "subgraphs" like these ones given by Fig. 2.

\begin{figure}[!h]
\begin{center}
\includegraphics[angle=0,scale=0.70]{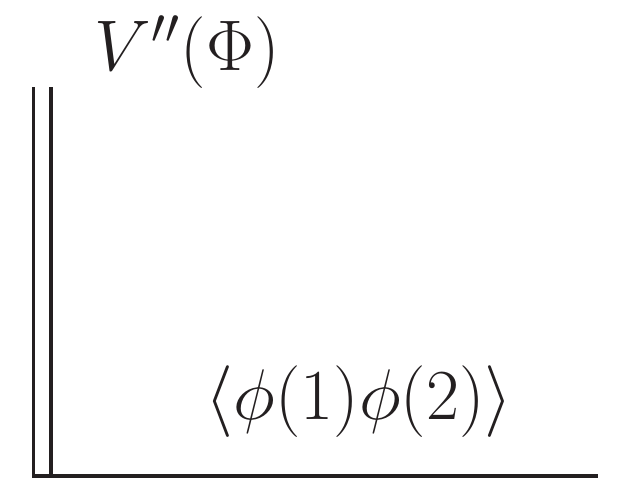}
\end{center}
\caption{A typical vertex in one-loop supergraphs.}
\end{figure}

Hence, the contribution of this subgraph is given by
\bea
Q_{12}=-\frac{V^{\prime\prime}_1}{g(k^2)+f(k^2)D_1^2}\delta_{12} \ ,
\eea
where $V^{\prime\prime}_1\equiv V^{\prime\prime}(\Phi(p_1=0,\theta_1))$. It follows from the result above that the contribution of a supergraph formed by $n$ subgraphs is given by
\bea
I_n&=&\int d^3x\frac{1}{2n}\int d^2\theta_1d^2\theta_2\ldots d^2\theta_{n}\int \frac{d^3k}{(2\pi)^3}Q_{12}Q_{23}\ldots Q_{n-1,n}Q_{n,1} \nonumber\\
&=&\int d^3x\frac{1}{2n}\int d^2\theta_1d^2\theta_2\ldots d^2\theta_{n}\int \frac{d^3k}{(2\pi)^3}\Big[-\frac{V^{\prime\prime}_1}{g(k^2)+f(k^2)D_1^2}\delta_{12}\Big]\nonumber\\
&\times&\Big[-\frac{V^{\prime\prime}_2}{g(k^2)+f(k^2)D_2^2}\delta_{23}\Big]\ldots\Big[-\frac{V^{\prime\prime}_n}{g(k^2)+f(k^2)D_n^2}\delta_{n,1}\Big] \ ,
\eea
where $2n$ is a symmetry factor. Such a factor takes into account the Taylor series expansion coefficients of the effective action, the usual symmetry factor of each supergraph, and the number of topologically distinct supergraphs \cite{Hat}.

We can integrate by parts the expression $I_n$ and discard terms involving covariant derivatives of $\Phi$ to get
\bea
\label{supergraphs}
I_n=\int d^5z\int \frac{d^3k}{(2\pi)^3}\frac{1}{2n}(-V^{\prime\prime})^n\bigg[\frac{1}{g(k^2)+f(k^2)D^2}\bigg]^{n}\delta_{\theta\theta^{\prime}}|_{\theta=\theta^{\prime}}.
\eea
The one-loop correction to the effective action is given by the sum of all supergraphs $I_n$,
\bea
\label{supergraphsEA}
\Gamma^{(1)}[\Phi]=\sum_{n=1}^{\infty}I_n=\int d^5z\int \frac{d^3k}{(2\pi)^3}\sum_{n=1}^{\infty}\frac{1}{2n}(-V^{\prime\prime})^n\bigg[\frac{1}{g(k^2)+f(k^2)D^2}\bigg]^{n}\delta_{\theta\theta^{\prime}}|_{\theta=\theta^{\prime}} \ .
\eea
This expression is rather generic. At this stage of the calculation, we have to specify the operator $\hat R$  in order to proceed with the calculation of $\Gamma^{(1)}$. The result of the complete evaluation of the D-algebra essentially depends on the explicit form of the operator $\hat{R}$. So, let us consider three characteristic examples where the final result is expressed in closed form and in terms of elementary functions.

As our first example, let us take $f=0$ and $g\ne0$. It follows from (\ref{supergraphsEA})
\bea
\Gamma^{(1)}[\Phi]=\int d^5z\int \frac{d^3k}{(2\pi)^3}\sum_{n=1}^{\infty}\frac{1}{2n}\bigg[-\frac{V^{\prime\prime}}{g(k^2)}\bigg]^{n}\delta_{\theta\theta^{\prime}}|_{\theta=\theta^{\prime}} \ .
\eea
Therefore, due to the property of the Grassmann delta function $\delta_{\theta\theta^{\prime}}|_{\theta=\theta^{\prime}}=0$, the KEP is given by
\bea
\label{result1}
K^{(1)}(\Phi)=0 \ ,
\eea
for $\hat R=g(\Box)$. In fact, we can make a stronger statement. From (\ref{higher1}), the model under consideration is given by
\bea
\label{model1}
S_{HR}=\frac{1}{2}\int d^5z\Phi g(\Box)\Phi+\int d^5zV(\Phi) \ .
\eea
Notice that there is no spinor derivative in the model (\ref{model1}). Hence, there will not be any spinor derivative neither in the propagators nor in the vertices. It follows from the supergraph rules that in an arbitrary $n$-loop contribution to the effective action, all delta functions $\delta(\theta_i-\theta_{i+1})$ coming from the propagator can be used to trivially evaluate the $d^2\theta_i$ integrals coming from the vertices. As a result, the $n$-loop correction for the effective action will have the final structure
\begin{align}
\Gamma^{(n)}[\Phi]=\sum_n\int d^3x_1\ldots d^3x_n\int d^2\theta{\cal T}(x_1,\ldots&,x_n)F_1(\Phi(x_1,\theta))\ldots F_n(\Phi(x_n,\theta))\delta_{\theta\theta^{\prime}}|_{\theta=\theta^{\prime}}\nonumber \\
\Rightarrow \Gamma^{(n)}[\Phi]&=0 \ .
\end{align}
Thus, from the result above and (\ref{loopexpansionEA}) we get
\bea
\Gamma[\Phi]=S_{HR}[\Phi] \ .
\eea
Therefore, the theory (\ref{model1}) is finite and does not have quantum corrections. This result is already known in four dimensions \cite{BuKu}. However, we have explicitly demonstrated its manifestation in three dimensions.

Our second example is $f=\xi(-\Box)^m$, and $g=0$, where $\xi$ is a parameter with a non-trivial mass dimension $[\xi]=[M]^{-2m}$, $\xi>0$, and $m$ is a non-negative integer. Consequently, we have
\bea
\Gamma^{(1)}[\Phi]&=&\int d^5z\int \frac{d^3k}{(2\pi)^3}\sum_{n=1}^{\infty}\frac{1}{2n}(-V^{\prime\prime})^n\bigg[\frac{1}{\xi(k^2)^mD^2}\bigg]^{n}\delta_{\theta\theta^{\prime}}|_{\theta=\theta^{\prime}}\nonumber\\
&=&\int d^5z\int \frac{d^3k}{(2\pi)^3}\sum_{n=1}^{\infty}\frac{1}{2n}\bigg[\frac{V^{\prime\prime}}{\xi(k^2)^{m+1}}\bigg]^n(D^2)^n\delta_{\theta\theta^{\prime}}|_{\theta=\theta^{\prime}} \ .
\eea
It can be shown that $(D^2)^n\delta_{\theta\theta^{\prime}}|_{\theta=\theta^{\prime}}=0$, for $n=2l$; $(D^2)^n\delta_{\theta\theta^{\prime}}|_{\theta=\theta^{\prime}}=(\sqrt{-k^2})^{n-1}$, for $n=2l+1$; where $l$ is an integer non-negative. Hence,  we get
\bea
\Gamma^{(1)}[\Phi]&=&\int d^5z\int \frac{d^3k}{(2\pi)^3}\frac{1}{2\sqrt{-k^2}}\sum_{l=0}^{\infty}\frac{1}{2l+1}\bigg[\frac{V^{\prime\prime}\sqrt{-k^2}}{\xi(k^2)^{m+1}}\bigg]^{2l+1}\nonumber \\
&=&\frac{1}{2}\int d^5z\int \frac{d^3k}{(2\pi)^3}\frac{1}{|k|}\arctan\bigg[\frac{V^{\prime\prime}|k|}{\xi(k^2)^{m+1}}\bigg] \ ,
\eea
where we have used the fact that $\sqrt{-k^2}=i|k|$ and the identity $\arctan(x)=\frac{1}{i}\textrm{arctanh}(ix)$. The integral above was solved in \cite{IYouHe}. Then we obtain
\bea
\label{result2EA}
\Gamma^{(1)}[\Phi]=\int d^5z\frac{1}{16\pi}\sec\bigg(\frac{\pi}{2m+1}\bigg)\bigg[\frac{V^{\prime\prime}(\Phi)}{\xi}\bigg]^{\frac{2}{2m+1}} \ .
\eea
The complete one-loop KEP can be read off from (\ref{result2EA}). As a result, we trivially obtain
\bea
\label{result2}
K^{(1)}(\Phi)=\frac{1}{16\pi}\sec\bigg(\frac{\pi}{2m+1}\bigg)\bigg[\frac{V^{\prime\prime}(\Phi)}{\xi}\bigg]^{\frac{2}{2m+1}} \ ,
\eea
for $\hat R=\xi(-\Box)^mD^2$, $m=0,1,2,\ldots$ \ . We notice that the one-loop correction for the KEP is finite and its finitenes is independent of the form of the potential $V(\Phi)$, as well as the KEP in the three-dimensional theory without higher derivatives \cite{ourEP}.  Moreover, the result (\ref{result2}) is highly generic. In particular, if $m=0$ and $\xi=1$, we get
\bea
K^{(1)}(\Phi)=-\frac{1}{16\pi}[V^{\prime\prime}(\Phi)]^2 \ ,
\eea
This is the (Euclidean) one-loop KEP for the three-dimensional real scalar superfield theory without higher derivatives. Our result is in agreement with the one obtained in \cite{ourEP}.

 Our last example will be $f=\xi_f(-\Box)^l$ and $g=\xi_g(-\Box)^{2l+1}$, where $l$ is a positive integer,  $\xi_f$ and $\xi_g$ are positive parameters with a non-trivial mass dimension $[\xi_f]=[M]^{-2l}$ and $[\xi_g]=[M]^{-2(2l+1)}$, respectively. It follows that we can rewrite (\ref{supergraphsEA}) as
\bea
\label{eq24}
\Gamma^{(1)}[\Phi]&=&\int d^5z\int \frac{d^3k}{(2\pi)^3}\sum_{n=1}^{\infty}\frac{1}{2n}(-V^{\prime\prime})^n\bigg[\frac{1}{\xi_g(k^2)^{2l+1}+\xi_f(k^2)^lD^2}\bigg]^{n}
\delta_{\theta\theta^{\prime}}|_{\theta=\theta^{\prime}}\nonumber\\
&=&\int d^5z\int \frac{d^3k}{(2\pi)^3}\sum_{n=1}^{\infty}\frac{1}{2n}(-V^{\prime\prime})^n\frac{1}{(n-1)!}\int_{0}^{\infty}ds \ s^{n-1}\nonumber\\
&\times&\exp[-s(\xi_g(k^2)^{2l+1}+\xi_f(k^2)^lD^2)]\delta_{\theta\theta^{\prime}}|_{\theta=\theta^{\prime}} \ ,
\eea
where we have used the Schwinger-DeWitt representation \cite{BaVi}
\bea
\frac{1}{\hat O^n}=\frac{1}{(n-1)!}\int_{0}^{\infty}ds \ s^{n-1}e^{-s\hat O} \ .
\eea
We can sum over $n$ in order to rewrite (\ref{eq24}) as
\bea
\Gamma^{(1)}[\Phi]&=&\int d^5z\int_{0}^{\infty}ds \ \frac{1}{2s}(-2+e^{-sV^{\prime\prime}})\int \frac{d^3k}{(2\pi)^3}e^{-s\xi_g(k^2)^{2l+1}}\sum_{m=0}^{\infty}\frac{(-s\xi_f(k^2)^l)^m}{m!}\nonumber\\
&\times&(D^2)^m\delta_{\theta\theta^{\prime}}|_{\theta=\theta^{\prime}} \ ,
\eea
where we have expanded the exponential argument. We can eliminate terms which does not depend on the background superfield by means of the normalization of the effective action. Hence, we obtain
\bea
\Gamma^{(1)}[\Phi]=\int d^5z\int_{0}^{\infty}ds \ \frac{1}{2s}e^{-sV^{\prime\prime}}\int \frac{d^3k}{(2\pi)^3}e^{-s\xi_g(k^2)^{2l+1}}\frac{1}{\sqrt{-k^2}}\sum_{n=0}^{\infty}\frac{(-s\xi_f(k^2)^l\sqrt{-k^2})^{2n+1}}{(2n+1)!} \ ,
\eea
where we use the previous result $(D^2)^m\delta_{\theta\theta^{\prime}}|_{\theta=\theta^{\prime}}=0$, for $m=2n$; $(D^2)^m\delta_{\theta\theta^{\prime}}|_{\theta=\theta^{\prime}}=(\sqrt{-k^2})^{m-1}$, for $m=2n+1$; where $n$ is an integer non-negative. Summing over $n$, we get
\bea
\Gamma^{(1)}[\Phi]=-\frac{1}{2}\int d^5z\int_{0}^{\infty}ds \ \frac{1}{s}e^{-sV^{\prime\prime}}\int \frac{d^3k}{(2\pi)^3}\frac{1}{|k|}e^{-s\xi_g(k^2)^{2l+1}}\sin\big[s\xi_f(k^2)^l|k|\big] \ .
\eea
The two integrals above can be solved by induction. As a result,
\bea
\label{result3EA}
\Gamma^{(1)}[\Phi]=\int d^5z-\frac{1}{8\pi}\sec\bigg(\frac{\pi}{2l+1}\bigg)\bigg[\frac{V^{\prime\prime}(\Phi)}{\xi_g}\bigg]^{\frac{1}{2l+1}}
\textrm{arcsinh}\bigg\{\frac{2}{2l+1}\sinh\bigg[\frac{\xi_f}{2\sqrt{\xi_gV^{\prime\prime}(\Phi)}}\bigg]\bigg\} \ .
\eea
Again, the complete one-loop KEP can be read off from the Eq. (\ref{result3EA}). Therefore, we finally obtain
\bea
\label{result3}
K^{(1)}(\Phi)=-\frac{1}{8\pi}\sec\bigg(\frac{\pi}{2l+1}\bigg)\bigg[\frac{V^{\prime\prime}(\Phi)}{\xi_g}\bigg]^{\frac{1}{2l+1}}
\textrm{arcsinh}\bigg\{\frac{2}{2l+1}\sinh\bigg[\frac{\xi_f}{2\sqrt{\xi_gV^{\prime\prime}(\Phi)}}\bigg]\bigg\} \ ,
\eea
for $\hat R=\xi_g(-\Box)^{2l+1}+\xi_f(-\Box)^lD^2$, $l=1,2,3,\ldots$ \ . We notice that this result, as well as the previous ones, is finite and does not need any renormalization  independently of the form of the potential $V(\Phi)$.

Let us move on to the calculation of the KEP for higher-derivative models involving the complex scalar superfield. Since the general idea
of the calculation is quite similar to the one described above, we will not go through all details.

Similarly to (\ref{realfield}), the three-dimensional complex superfield theory is described by the action
\bea
\label{complexfield}
S_{C}=\int d^5z\big[\bar\Phi D^2\Phi+V(\bar\Phi\Phi)\big] \ .
\eea
However, differently from (\ref{realfield}), this theory has a global symmetry. The action (\ref{complexfield}) is invariant under the following global transformation: $\Phi\to e^{iK}\Phi$. Therefore, in order to introduce higher derivatives, we will demand that the higher-derivative terms do not explicitly break the symmetry of the theory (\ref{complexfield}). Again, for simplicity, we also demand that the new action does not contain interaction terms with higher derivatives. Consequently, we will consider the theory described by the action:
\bea
\label{higher2}
S_{HC}=\int d^5z\big[\bar\Phi\hat R\Phi+V(\bar\Phi\Phi)\big] \ ,
\eea
where $\hat R=g(\Box)+f(\Box)D^2$.

We can expand (\ref{higher2}) around the background superfields $\Phi+\sqrt{\hbar}\phi$, $\bar\Phi+\sqrt{\hbar}\bar\phi$, and keep only the quadratic terms in quantum superfields. Hence, we get
\bea
\label{complexquadraticfields}
S_2[\Phi,\bar\Phi;\phi,\bar\phi]=\int d^5z\big[\bar\phi(g(\Box)+f(\Box)D^2)\phi+V_{\Phi\bar\Phi}\phi\bar\phi+\frac{1}{2}V_{\Phi\Phi}\phi^2+\frac{1}{2}V_{\bar\Phi\bar\Phi}\bar\phi^2\big] \ ,
\eea
where we have used a shorthand notation $V_{\bar\Phi\Phi}=\frac{\partial^2V(\bar\Phi,\Phi)}{\partial\bar\Phi\partial\Phi}$, $V_{\Phi\Phi}=\frac{\partial^2V(\bar\Phi,\Phi)}{\partial\Phi^2}$, $V_{\bar\Phi\bar\Phi}=\frac{\partial^2V(\bar\Phi,\Phi)}{\partial\bar\Phi^2}$. It is convenient to rewrite the functional (\ref{complexquadraticfields}) in a matrix form, namely
\bea
\label{quadraticaction2}
S_2[\Phi,\bar\Phi;\phi,\bar\phi]=\frac{1}{2}\int d^5z\big[\phi^i{P_i}^j(g(\Box)+f(\Box)D^2)\phi_j+\phi^i{M_i}^j\phi_j\big] \ .
\eea
where
\begin{equation}
\phi_i=\left(\begin{array}{c}
\phi\\
\bar\phi
\end{array}\right)
 \ , \ \phi^i=\left(\begin{array}{cc}
\phi \ & \ \bar\phi
\end{array}\right)
\ , \ {P_i}^j=\left(\begin{array}{cc}
0 & 1\\
1 & 0
\end{array}\right)
\ , \ {M_i}^j=\left(\begin{array}{cc}
V_{\Phi\Phi} & V_{\Phi\bar\Phi}\\
V_{\Phi\bar\Phi} & V_{\bar\Phi\bar\Phi}
\end{array}\right) \ .
\end{equation}
It follows that the propagator is given by
\bea
\label{propagator2}
\langle \phi_i(1)\phi^j(2)\rangle=-\frac{{P_i}^j}{g(k^2)+f(k^2)D_1^2}\delta_{12} \ .
\eea
The one-loop supergraphs will have the same pattern as the ones in the Fig. 1, except by the fact that now each supergraph will be formed by $n$ subgraphs like these ones depicted at Fig. 3.
\begin{figure}[!h]
\begin{center}
\includegraphics[angle=0,scale=0.70]{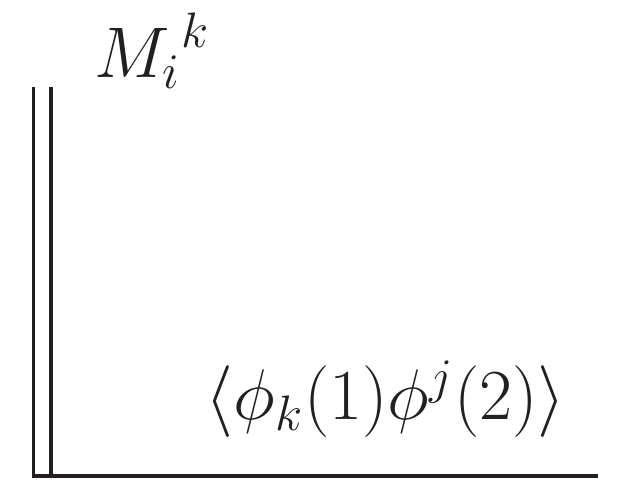}
\end{center}
\caption{A matrix vertex in one-loop supergraphs.}
\end{figure}

Hence, the contribution of this subgraph is given by
\bea
{(Q_{12})_i}^j&=&-{(M_1)_i}^k{P_k}^j\frac{1}{g(k^2)+f(k^2)D_1^2}\delta_{12}=-{(\widetilde M_1)_i}^j\frac{1}{g(k^2)+f(k^2)D_1^2}\delta_{12} \ ,\\
\widetilde M&=&\left(\begin{array}{cc}
V_{\Phi\bar\Phi} & V_{\Phi\Phi}\\
V_{\bar\Phi\bar\Phi} & V_{\Phi\bar\Phi}
\end{array}\right) \ .
\eea
It follows from the result above that the contribution of a supergraph formed by $n$ subgraphs is given by
\bea
J_n&=&\int d^3x\frac{1}{2n}\int d^2\theta_1d^2\theta_2\ldots d^2\theta_{n}\int \frac{d^3k}{(2\pi)^3}\textrm{Tr}\{{(Q_{12})_i}^j{(Q_{23})_j}^k\ldots {(Q_{n-1,n})_l}^m{(Q_{n,1})_m}^p\} \nonumber\\
&=&\int d^3x\frac{1}{2n}\int d^2\theta_1d^2\theta_2\ldots d^2\theta_{n}\int \frac{d^3k}{(2\pi)^3}\textrm{Tr}\big\{\Big[-{(\widetilde M_1)_i}^j\frac{1}{g(k^2)+f(k^2)D_1^2}\delta_{12}\Big]\nonumber\\
&\times&\Big[-{(\widetilde M_2)_j}^k\frac{1}{g(k^2)+f(k^2)D_2^2}\delta_{23}\Big]\ldots\Big[-{(\widetilde M_n)_m}^p\frac{1}{g(k^2)+f(k^2)D_n^2}\delta_{n,1}\Big]\big\} \ ,
\eea

After successive integration by parts and summing all supergraphs $J_n$, we get the effective action
\bea
\Gamma^{(1)}[\Phi]=\int d^5z\int \frac{d^3k}{(2\pi)^3}\sum_{n=1}^{\infty}\frac{1}{2n}\textrm{Tr}[\widetilde M^n]\bigg[\frac{-1}{g(k^2)+f(k^2)D^2}\bigg]^{n}\delta_{\theta\theta^{\prime}}|_{\theta=\theta^{\prime}} \ .
\eea
The trace of the matrix $\widetilde M^n$ can be calculated by means of the $\widetilde M$'s eigenvalues, which are $\lambda_{1,2}=V_{\bar\Phi\Phi}\pm(V_{\Phi\Phi}V_{\bar\Phi\bar\Phi})^{1/2}$. Therefore, $\textrm{Tr}[\widetilde M^n]=\lambda_1^n+\lambda_2^n$. It follows that
\bea
\label{complexsupergraphsEA}
\Gamma^{(1)}[\Phi]&=&\int d^5z\int \frac{d^3k}{(2\pi)^3}\sum_{n=1}^{\infty}\frac{1}{2n}(-\lambda_1)^n\bigg[\frac{1}{g(k^2)+f(k^2)D^2}\bigg]^{n}\delta_{\theta\theta^{\prime}}|_{\theta=\theta^{\prime}}\nonumber\\
&+&\int d^5z\int \frac{d^3k}{(2\pi)^3}\sum_{n=1}^{\infty}\frac{1}{2n}(-\lambda_2)^n\bigg[\frac{1}{g(k^2)+f(k^2)D^2}\bigg]^{n}\delta_{\theta\theta^{\prime}}|_{\theta=\theta^{\prime}} \ .
\eea
We notice that the expression (\ref{complexsupergraphsEA}) is quite analogous to (\ref{supergraphsEA}). Therefore, we will not need to reproduce the calculations.

The one-loop KEPs for the three-dimensional higher-derivative complex superfield theory are given by
\bea
\label{complexresult1}
K^{(1)}(\Phi)=0 \ ,
\eea
for $\hat R=g(\Box)$. Moreover, $\Gamma[\Phi]=S_{HC}[\Phi]$.
\bea
\label{complexresult2}
K^{(1)}(\Phi)&=&\frac{1}{16\pi}\sec\bigg(\frac{\pi}{2m+1}\bigg)\bigg\{\bigg[\frac{V_{\bar\Phi\Phi}+(V_{\Phi\Phi}V_{\bar\Phi\bar\Phi})^{1/2}}{\xi}\bigg]^{\frac{2}{2m+1}}\nonumber\\
&+&\bigg[\frac{V_{\bar\Phi\Phi}-(V_{\Phi\Phi}V_{\bar\Phi\bar\Phi})^{1/2}}{\xi}\bigg]^{\frac{2}{2m+1}}\bigg\} \ ,
\eea
for $\hat R=\xi(-\Box)^mD^2$, $m=0,1,2,\ldots$ \ .
\bea
\label{complexresult3}
K^{(1)}(\Phi)&=&-\frac{1}{8\pi}\sec\bigg(\frac{\pi}{2l+1}\bigg)\bigg\{\bigg[\frac{V_{\bar\Phi\Phi}+(V_{\Phi\Phi}V_{\bar\Phi\bar\Phi})^{1/2}}{\xi_g}\bigg]^{\frac{1}{2l+1}}\nonumber\\
&\times&\textrm{arcsinh}\bigg\{\frac{2}{2l+1}\sinh\bigg[\frac{\xi_f}{2\sqrt{\xi_g(V_{\bar\Phi\Phi}+(V_{\Phi\Phi}V_{\bar\Phi\bar\Phi})^{1/2})}}\bigg]\bigg\}+
\bigg[\frac{V_{\bar\Phi\Phi}-(V_{\Phi\Phi}V_{\bar\Phi\bar\Phi})^{1/2}}{\xi_g}\bigg]^{\frac{1}{2l+1}}\nonumber\\
&\times&\textrm{arcsinh}\bigg\{\frac{2}{2l+1}\sinh\bigg[\frac{\xi_f}{2\sqrt{\xi_g(V_{\bar\Phi\Phi}-(V_{\Phi\Phi}V_{\bar\Phi\bar\Phi})^{1/2})}}\bigg]\bigg\}\bigg\} \ ,
\eea
for $\hat R=\xi_g(-\Box)^{2l+1}+\xi_f(-\Box)^lD^2$, $l=1,2,3,\ldots$ \ .

It is worth noticing that our results are finite and does not need any renormalization independently of the form of the potential $V(\bar\Phi\Phi)$. Moreover, if we take in (\ref{complexresult2}) the particular values $m=0$, $\xi=1$, and $V(\bar\Phi,\Phi)=\frac{\lambda}{2}(\bar\Phi\Phi)^2$, we recover the (Euclidean) one-loop KEP for the three-dimensional complex scalar superfield theory without higher derivatives, whose result was originally obtained in \cite{WHD} in the context of the gauge theory.

We succeeded in calculation of the one-loop effective potential for the three-dimensional higher-derivative scalar superfield theories, both for real and complex superfields. The result was found to be finite, first of all, due to the presence of higher derivatives (we note that the higher-derivative supersymmetric theory in $3D$ is all-loop finite). At the same time, one should note that we treat this theory as an effective one, without addressing the issue of ghost states whose presence is characteristic for the higher-derivative theories \cite{Ant}. We hope to study the impact of ghosts in a forthcoming paper.

\vspace{5mm}

{\bf Acknowledgments.}
This work was partially supported by Conselho Nacional de
Desenvolvimento Cient\'\i fico e Tecnol\'ogico (CNPq). A. Yu. P. has
been supported by the CNPq project No. 303438-2012/6. The work by F. S. Gama has been supported by the
CNPq process No. 141228/2011-3.

\end{document}